\documentclass[onecolumn,floatfix,pra,superscriptaddress,nofootinbib]{revtex4}
\usepackage{color}
\usepackage{graphicx}
\usepackage{bm}
\usepackage{amsmath}
\usepackage{enumerate}

\newcommand{\cl}{ \text{cl} }
\newcommand{\ti}{ \tilde }
\newcommand{\ep}{ \epsilon }
\newcommand{\pa}{ \partial }
\begin{document}
\title{ Dissipative tunnelling by means of scaled trajectories }
\author{S. V. Mousavi}
\email{vmousavi@qom.ac.ir}
\affiliation{Department of Physics, University of Qom, Ghadir Blvd., Qom 371614-6611, Iran}
\author{S. Miret-Art\'es}
\email{s.miret@iff.csic.es}
\affiliation{Instituto de F\'isica Fundamental, Consejo Superior de
Investigaciones Cient\'ificas, Serrano 123, 28006 Madrid, Spain}
\begin{abstract}
Dissipative quantum tunnelling through an inverted parabolic barrier is considered in the 
presence of an electric field. 
​A Schr\"{o}dinger-Langevin or Kostin quantum-classical transition wave equation is used 
and applied resulting in a scaled differential equation of motion.
A Gaussian wave packet solution to the resulting scaled Kostin nonlinear equation is assumed 
and compared to the same solution for the scaled linear Caldirola-Kanai  equation. The resulting
scaled trajectories are obtained at different dynamical regimes and friction cases, showing
the gradual decoherence process in this open dynamics.
Theoretical results show that the transmission probabilities are always higher in  the Kostin  approach
than in the Caldirola-Kanai approach in the presence or not of an external electric field. This
discrepancy should be understood due to the presence of an environment since the corresponding
open dynamics should be governed by nonlinear quantum equations, whereas the second
approach is issued from an effective Hamiltonian within a linear theory.
%
%These theoretical results may be checked by experiment. Thus, experiment can decide between 
%two models. 
%
\end{abstract}
\maketitle
%========================================
%========================================
\section{Introduction}

Dissipative tunnelling in the presence or not of an electric field has many applications in 
transport properties, reactive scattering, quantum optics, molecular biology, etc. One 
of the main goals is to analyze the gradual decoherence process existing in this particular dynamics
by using very different theoretical methods within the density matrix, path-integral and 
Langevin formalisms \cite{Weiss-book-1999,Razavy-book-2005}. A new alternative way which is 
by far much less used is the Bohmian formalism \cite{NaMi-book-2017} where the decoherence
process is described in terms of trajectories leading, in our opinion, to a more intutitive way 
of understanding it. In this sense, dealing with analytically solvable models is very useful in order 
to gain new insights.

By considering dissipation from a phenomenological way, the tunnelling dynamics by an inverted
parabollic barrier is very convenient because it provides all the main ingredients to tackle with
success  such an endeavour as well as to compare with existing results coming from different 
theoretical treatments. In particular, comparison with the works by Baskoutas and Jannussis 
\cite{BaJa-JPA-1992} and Papadopoulos \cite{Pa-JPA-1997}.  
Our purpose is to analyze two different approaches to this dissipative dynamics,  the nonlinear, logarithmic
Schr\"odinger-Langevin (or Kostin) equation \cite{Kostin-1972} and the linear Schr\"odinger equation 
coming from the so-called Cardirola-Kanai Hamiltonian  \cite{caldirola,kanai,SaCaRoLoMi-AP-2014,
MoMi-JPCO-2018} within the Bohmian formalism \cite{Holland-book-1993,salva1,salva2}.  
Recently, Tokieda and Hagino \cite{ToHa-PRC-2017} have considered the same approaches to study dissipative
tunneling without the presence of an  electrical field by solving directly the corresponding wave equations.
The gradual decoherence process is better studied by using the so-called quantum-classical 
transition wave equation, originally proposed by Richardson et al. \cite{RiSchMaVaBa-PRA-2014} 
in the context of conservative systems. This quantum-classical transition is governed by a 
continuous parameter covering these two regimes as being the two extreme cases. Recently, Chou has applied 
this wave equation to analyze wave-packet  interference \cite{Chou1-2016}  and the dynamics of the harmonic 
and Morse oscillators with complex trajectories \cite{Chou2-2016}. Here, we have extended this 
procedure to dissipative quantum dynamics.
Doing this, we have a wave equation even in the classical regime and the Born rule is assumed in 
this regime too. Then, by considering the {\it actual momentum} \cite{Holland-book-1993}
distribution function of particles in the classical ensemble, 
%we will see initially all particles have the same momentum when the classical wave has a Gaussian form in the coordinate %space. In this way 
we have a strict answer to the question of classical phase space distribution function 
which is problematic otherwise \cite{HoPaBa-JPA-2009}. 
The resulting trajectories have been called {\it scaled trajectories} 
\cite{MoMi-JPCO-2018} since a scaled Planck's constant in terms of that parameter is used. 
Furthermore, by assuming a time-dependent Gaussian ansatz for the probability density,
theses scaled trajectories are written as a sum of  a classical trajectory (a particle property) 
plus a term containing the width of the corresponding wave packet (a wave property) within 
of what has been called {\it dressing scheme} \cite{NaMi-book-2017}. In the quantum regime,
the corresponding trajectories are the well-known quantum trajectories due to Bohm which display
the noncrossing property which, in general, is no longer valid in the classical regime but 
it is kept in the transtion regime. 
However, in this work, this property is still valid in the classical regime by construction of the
transition wave equation itself. This new aspect together with the Born rule for the distribution of 
particles' position in the classical ensemble lead us to have a good criterium for tunnelling. 

The procedure of using a continuous parameter monitoring the different regimes in the 
theory (a scaled Planck's constant) is apparently quite similar to the WKB approach (a series expansion
in powers of Planck's constant), widely used in conservative systems. However, there are several important
differences. First, in the WKB,  the classical Hamilton-Jacobi equation for the classical action at zero order
of the expansion in powers of $\hbar$ is obtained whereas, in the scaling procedure, the so-called 
classical wave (nonlinear) equation \cite{Schi-PR-1962} is reached by construction. Second, the hierarchy of the
differential equations for the action at different orders of the expansion in $\hbar$ is substituted by only a
transition wave equation which can be easily solved in the linear domain. Third, in the Bohmian 
framework, the transition from quantum to classical trajectories is carried out in a continuous way allowing 
us to follow the continuity of the trajectories when changing of regime. Fourth, the scaling procedure 
extended and applied to open (dissipative and/or stochastic) quantum systems is very easy to implement.
And fifth, the gradual decoherence process due to the scaled Planck's constant can be seen as an extra source
which it has to be added to the decoherence  due to the presence of an environment. 

The organization of this work is as follows. In Section II, the nonlinear (logarithmic) 
Schr\"odinger-Langevin or Kostin equation within the context of open quantum systems 
as well as the corresponding scaled equation are introduced. They are then specialized to 
the dissipative case. In Section III, the Cardirola-Kanai formalism, where dissipation 
is introduced from a phenomenological point of view, is developed leading to a linear Schr\"odinger
equation. Section IV deals with the dissipative Bohmian dynamics and scaled trajectories by 
assuming a time-dependent Gaussian ansatz for the probability density. The presence of the field 
in this dissipative tunnelling dynamics is considered in Section V. Finally, results and discussion as 
well as some conclusions are presented in the remaining two sections.

%=====================================================
%=====================================================
\section{Quantum-classical transition and scaled Schr\"{o}dinger-Langevin equations} \label{sec: Sch-La we}

Kostin derived heuristically from the standard Langevin equation the so-called Schr\"{o}dinger-Langevin or Kostin nonlinear (logarithmic) equation which is written in one dimension as 
\cite{Kostin-1972,NaMi-book-2017} 
\begin{eqnarray} 
i \hbar \frac{\partial}{\partial t}\psi(x, t) &=& \left[ -\frac{\hbar^2}{2m} \frac{\partial^2}{\partial x^2}
+ V(x, t) + V_r(x, t) + \frac{\gamma \hbar}{2 i} \left( \ln \frac{\psi}{\psi^*} - \left \langle \ln \frac{\psi}{\psi^*} \right \rangle \right) 
\right] \psi(x, t) ,
\label{eq: Sch_Lan}
\end{eqnarray}
where $m$ is the mass of the quantum particle, $\gamma$ the friction coefficient, $V(x)$ is 
the interaction potential and $V_r(x)$ the random potential given by
\begin{eqnarray} \label{eq: random_pot}
V_r(x, t) &=& x~F_r(t) ,
\end{eqnarray}
$F_r(t)$ being a time-dependent random force. 
%Without this random force, Eq. (\ref{eq: Sch_Lan}) is just Kostin equation.
%
Following \cite{RiSchMaVaBa-PRA-2014},  
Eq. (\ref{eq: Sch_Lan}) can be rewritten as a quantum-classical transition wave equation  as follows 
\begin{eqnarray} 
i \hbar \frac{\partial}{\partial t}\psi_{\ep}(x, t) &=& \bigg[ -\frac{\hbar^2}{2m}
\frac{\partial^2}{\partial x^2} + V(x, t) + V_r(x, t) 
+ \frac{\gamma \hbar}{2 i} \left( \ln \frac{\psi_{\ep}}{\psi_{\ep}^*} - \left \langle \ln \frac{\psi_{\ep}}{\psi_{\ep}^*} \right \rangle \right) 
\nonumber \\
& &~~+ (1-\ep) \frac{\hbar^2}{2m} 
\frac{1}{|\psi_{\epsilon}(x, t)|} 
\frac{\partial^2 |\psi_{\epsilon}(x, t)|}{\partial x^2 } 
\bigg ] \psi_{\ep}(x, t) ,  \label{eq: quan_class transition}
\end{eqnarray}
where a degree of quantumness given by the $\ep$ parameter, with $ 0 \leq \ep \leq 1 $, is included 
by means of an extra sum representing the quantum potential of the corresponding Bohmian dynamics 
\cite{Holland-book-1993,NaMi-book-2017},
\begin{eqnarray} \label{qpe}
Q_{\ep}(x, t) &=& - \frac{\hbar^2}{2m} \frac{1}{|\psi_{\ep}(x, t)|} 
\frac{\partial^2 |\psi_{\ep}(x, t)|}{\partial x^2 }   .
\end{eqnarray}
This equation provides a {\it continuous} or gradual description for the transition process of physical
systems from purely quantum, $\ep=1$, to purely classical, $\ep=0$, regime which is ruled by the
so-called classical wave equation \cite{Schi-PR-1962}. This parameter can also be 
seen as one defining the dynamical regime. The wave function 
$\psi_{\ep}(x, t)$ is thus affected by this parameter determining this transition or decoherence proces.
By substituting now the polar form of this wave function
\begin{eqnarray} \label{eq: tran_wf_polar}
\psi_{\epsilon}(x, t) &=& R_{\ep}(x, t) e^{i S_{\ep}(x, t)/ \hbar } 
\end{eqnarray}
into the transition wave equation (\ref{eq: quan_class transition}), the following coupled equations
for the amplitude $R_{\ep}(x,t)$ and phase $S_{\ep}(x, t)$  are obtained 
\begin{eqnarray}
\frac{\partial  R_{\ep}}{\partial t} &=&  -\frac{1}{2m } \left( 2 \frac{\partial  R_{\ep}}{\partial x} \frac{\partial  S_{\ep}}{\partial x} + R_{\ep} \frac{\partial^2  S_{\ep}}{\partial x^2} \right) , \label{eq: tran_imag_part}
\\
- \frac{\partial  S_{\ep}}{\partial t} R_{\ep} &=& - \frac{\hbar^2}{2m} \left[  \ep \frac{\partial^2  R_{\ep}}{\partial x^2}
- \frac{1}{\hbar^2} R_{\ep} \left( \frac{\partial  S_{\ep}}{\partial x} \right)^2
\right ] + \bigg[ V(x, t) + V_r(x, t) + \gamma ( S_{\ep} - \langle S_{\ep} \rangle  )  \bigg]  R_{\ep} ,
\label{eq: tran_real_part}
\end{eqnarray}
and where we have made use of the fact that
\begin{eqnarray} \label{eq: tran_wave_phase}
S_{\ep} &=& \frac{\hbar}{2i} \ln \frac{\psi_{\ep}}{\psi_{\ep}^*} .
\end{eqnarray}
Furthermore, by introducing the so-called scaled Plank's constant as
\begin{eqnarray} \label{eq: scaled Planck}
\ti{\hbar} &=& \hbar ~ \sqrt{\epsilon} ,
\end{eqnarray}
and the corresponding scaled wave function as
\begin{eqnarray} \label{eq: scaled_wf_polar}
\ti{\psi}(x, t) &=& R_{\ep}(x, t) e^{i S_{\ep}(x, t)/ \ti{\hbar} } ,
\end{eqnarray}
%
%with 
%
%\begin{eqnarray} \label{eq: wave_phase}
%S_{\ep} &=& \frac{\ti{\hbar}}{2i} \ln \frac{\ti{\psi}}{\ti{\psi}^*} 
%\end{eqnarray}
%
into Eqs. (\ref{eq: tran_imag_part}) and  (\ref{eq: tran_real_part}) and, after some 
straightforward algebraic manipulations, the following scaled nonlinear 
Schr\"{o}dinger-Langevin equation for a stochastic dynamics is again reached 
\begin{eqnarray} \label{eq: Scaled Sch-Lan-1}
i \ti{\hbar} \frac{\partial}{\partial t}\ti{\psi}(x, t) &=& \left[ -\frac{\ti{\hbar}^2}{2m} \frac{\partial^2}{\partial x^2}
+ V(x, t) + V_r(x, t) + \frac{\gamma \ti{\hbar}}{2 i} \left( \ln \frac{\ti{\psi}}{\ti{\psi}^*} - \left \langle \ln \frac{\ti{\psi}}{\ti{\psi}^*} \right \rangle \right) 
\right] \ti{\psi}(x, t)   ,
\end{eqnarray}
$\ti{\psi}(x, t)$ being the scaled wave function.
When the random potential $V_r(x,t)$ is neglected, the dissipative system is described by 
\begin{eqnarray} \label{eq: Scaled Sch-Lan}
i \ti{\hbar} \frac{\partial}{\partial t}\ti{\psi}(x, t) &=& \left[ -\frac{\ti{\hbar}^2}{2m} \frac{\partial^2}{\partial x^2}
+ V(x, t) +  \frac{\gamma \ti{\hbar}}{2 i} \left( \ln \frac{\ti{\psi}}{\ti{\psi}^*} - \left \langle \ln \frac{\ti{\psi}}{\ti{\psi}^*} \right \rangle \right) 
\right] \ti{\psi}(x, t)   ,
\end{eqnarray}
being again a nonlinear logarithmic equation, the scaled Kostin equation.
%and multiplying both sides of (\ref{eq: tran_imag_part}) by the factor $ i \ti{\hbar} 
%\exp[i S_{\ep}/\ti{\hbar} ]$ and both sides of (\ref{eq: tran_real_part}) by the exponential factor 
%$\exp[i S_{\ep}/\ti{\hbar} ]$ and finally adding both sides of resulting equations yields
%
%\begin{eqnarray}
%\text{LHS} &=& i \ti{\hbar} \left( \frac{\partial  R_{\ep}}{\partial t} e^{iS_{\ep}/\ti{\hbar}} + \frac{i}{\ti{\hbar}} \frac{\partial  S_{\ep}}{\partial t} \ti{\psi} \right) = i \ti{\hbar} \frac{\partial}{\partial t} 
%\left( R_{\ep} e^{iS_{\ep}/\ti{\hbar}} \right) ,
%\end{eqnarray}
%
%for the left hand side and 
%
%\begin{eqnarray}
%\text{RHS} &=& 
%-\frac{ \ti{\hbar} ^2 }{2m }  \left[
% \frac{\partial^2  R_{\ep}}{\partial x^2} 
%+ \frac{2i}{\ti{\hbar}} \frac{\partial  R_{\ep}}{\partial x} \frac{\partial  S_{\ep}}{\partial x} 
%+ \frac{i}{\ti{\hbar}} R_{\ep} \frac{\partial^2  S_{\ep}}{\partial x^2}   
%- \frac{1}{\ti{\hbar}^2} R_{\ep} \left( \frac{\partial  S_{\ep}}{\partial x} \right)^2 \right] e^{iS_{\ep}/\ti{\hbar}}
%\nonumber \\ 
%&+& \bigg[ V(x, t) + V_r(x, t) + \gamma ( S_{\ep} - \langle S_{\ep} \rangle  )  \bigg]  R_{\ep} e^{iS_{\ep}/\ti{\hbar}}
%\nonumber \\
%&=& 
%
%\left[
%-\frac{\ti{\hbar}^2}{2m } \frac{\partial^2 }{\partial x^2}
%+ V + V_r + \gamma ( S_{\ep} - \langle S_{\ep} \rangle  ) 
%\right] R_{\ep} e^{iS_{\ep}/\ti{\hbar}} ,
%\end{eqnarray}
%
%for the right hand side.
%
%
In any case, the transition wave equation (\ref{eq: quan_class transition}) 
is equivalent to the scaled nonlinear Sch\"odinger-Langevin equation  
(\ref{eq: Scaled Sch-Lan-1}) (or  (\ref{eq: Scaled Sch-Lan}) only for the dissipative case).  
Moreover, the corresponding wave functions and phases are related by
\begin{eqnarray} 
\ti{\psi}(x, t) &=& \psi_{\ep}(x, t) \exp \left[ \frac{i}{\hbar} \left( \frac{1}{\sqrt{\ep}} - 1  \right) S_{\ep}(x, t) \right],
\label{eq: scaled_transtion relation} 
%\\
%\ti{S}(x, t) &=& S_{\ep}(x, t) , \label{eq: scaled_transtion phase relation}
\end{eqnarray}
which are derived from Eqs. (\ref{eq: tran_wf_polar}) and (\ref{eq: scaled_wf_polar}).
%; and Eqs. (\ref{eq: tran_wave_phase}) and (\ref{eq: scaled_wave_phase}) respectively. 

Thus, the decoherence process resulting from the open quantum dynamics  and scaled Planck's
constant is carried out in a gradual way (it is worth mentioning that the environment is also seen as acting like a continuous measuring apparatus).

%=====================================================
%=====================================================

\section{Quantum-classical transition and scaled Schr\"{o}dinger equations in the CK approach}

The so-called classical CK Hamiltonian for dissipative systems is given by \cite{Razavy-book-2005}
\begin{equation} \label{eq: class_Ham}
H = \frac{p^2}{2m} e^{-\gamma t} + V(x) e^{\gamma t} 
\end{equation}
and its corresponding Hamiltonian operator $\hat{H}$ can be obtained from the standard 
quantization rule by substituting the momentum $p$ by 
$\frac{\hbar}{i} \frac{\partial}{\partial x}$,
\begin{eqnarray} \label{eq: KC QM-Hamiltonian}
\hat{H} &=& - \frac{\hbar^2}{2m} e^{-\gamma t} \frac{\partial^2}{\partial x^2} + 
e^{\gamma t} V(x) .
\end{eqnarray}
It is well known that the commutation relation  of the position and kinematic momentum operators 
is given by $[x,p]=i \hbar e^{-\gamma t}$, leading to the violation of the Heisenberg uncertainty
principle. By means of the transformation to the canonical variables ${\bar x}=x$ and ${\bar p}
= p e^{\gamma t}$, this principle is again fulfilled.  
Notice that as long as quantities related to the kinematic momentum are not
computed, the use of the corresponding wave equation in the coordinate space is formally correct
\cite{SaCaRoLoMi-AP-2014}. 
In this framework, friction shows the action of an effective (almost macroscopic) environment 
coupled to the particle making the motion more and more predictable (classical) as time proceeds.
Thus, the violation of the uncertainty principle has sometimes been justified from a dissipative
dynamics \cite{Pa-JPA-1997}.
The time-dependent Schr\"{o}dinger equation within the 
CK framework then reads  as
\begin{eqnarray} \label{eq: Sch_viscid}
i \hbar \frac{\partial}{\partial t}\psi(x, t) &=& \left[ -\frac{\hbar^2}{2m} e^{-\gamma t}
\frac{\partial^2}{\partial x^2} + e^{\gamma t}  V(x) 
\right] \psi(x, t) .
\end{eqnarray}
Following now the same procedure as in previous Section, a quantum-classical transition 
wave equation \cite{RiSchMaVaBa-PRA-2014} can again be introduced according to 
\cite{MoMi-JPCO-2018} as
\begin{eqnarray}
i \hbar \frac{\partial}{\partial t}\psi_{\epsilon}(x, t) &=& \left[ -\frac{\hbar^2}{2m} e^{-\gamma t}
\frac{\partial^2}{\partial x^2} + V(x) e^{\gamma t} + (1-\epsilon) 
\frac{\hbar^2}{2m} \frac{1}{|\psi_{\ep}(x, t)|} \frac{\partial^2 |\psi_{\ep}(x, t)|}{\partial x^2 }
e^{-\gamma t}
\right] \psi_{\epsilon}(x, t) ,  \label{eq: quan_class transition-ck}
\end{eqnarray}
This equation also provides a {\it continuous} or gradual description for the transition  or 
decoherence process of physical systems from purely quantum to classical regime in the 
CK framework. Then, by substituting the standard polar form of the wave function given by Eq. 
(\ref{eq: tran_wf_polar}) into Eq. (\ref{eq: quan_class transition-ck}) and after 
some straightforward manipulations, the following coupled equations are reached 
\begin{eqnarray}
- \frac{\partial  S_{\ep}}{\partial t} \ti{\psi} &=& \frac{1}{2m} e^{-\gamma t} \left( \frac{\partial  S_{\ep}}{\partial x} \right)^2  \ti{\psi} + V(x) e^{\gamma t} \ti{\psi} -  \frac{ \ti{\hbar}^2}{2m} e^{-\gamma t} \frac{1}{R_{\ep}}
\frac{\partial^2  R_{\ep}}{\partial x^2} \ti{\psi} ,  \label{eq: tran_real_part2}
\\
i \ti{\hbar} \frac{\partial  R_{\ep}}{\partial t} e^{iS_{\ep}/\ti{\hbar}} &=&  -\frac{\ti{\hbar}^2}{2m } e^{-\gamma t}  \left[ \frac{2i}{\ti{\hbar}}
\frac{\partial  R_{\ep}}{\partial x} \frac{\partial  S_{\ep}}{\partial x} e^{ iS_{\ep}/\ti{\hbar} } + \frac{i}{\ti{\hbar}} \frac{\partial^2  S_{\ep}}{\partial x^2} \ti{\psi} \right] , \label{eq: tran_imag_part2}
\end{eqnarray}
where the scaled Planck's constant defined in Eq. (\ref{eq: scaled Planck}) is used
with the scaled wave function in polar form written as Eq. (\ref{eq: scaled_wf_polar}).
By adding Eq. (\ref{eq: tran_real_part2}) and Eq. (\ref{eq: tran_imag_part2}),  the corresponding
scaled linear Schr\"{o}dinger equation
\begin{eqnarray} \label{eq: CK Scaled Sch}
i \tilde{\hbar} \frac{\partial}{\partial t}\ti{\psi}(x, t) &=& \left[ -\frac{\tilde{\hbar}^2}{2m} 
e^{-\gamma t} \frac{\partial^2}{\partial x^2} + V(x) e^{\gamma t} \right] \ti{\psi}(x, t) ,
\end{eqnarray}
is thus obtained in the CK framework.

Thus, the nonlinear transition equation (\ref{eq: quan_class transition-ck}) is equivalent to the 
scaled linear Schr\"{o}dinger equation (\ref{eq: CK Scaled Sch}) and has the same structure than 
Eq. (\ref{eq: Sch_viscid}). This will be our working equation for the scaled wave function, 
which can also be expressed in terms of the transition wave function after 
Eq. (\ref{eq: scaled_transtion relation}). As is clearly seen, the dissipative dynamics issued from 
this model can a priori be quite different from that provided by the nonlinear logaritmic 
Eq. (\ref{eq: Scaled Sch-Lan}).

%===============================================================
%==================================================================

\section{Bohmian dynamics of Gaussian wave packets. Scaled trajectories}

\subsection{The Schr\"{o}dinger-Langevin or Kostin approach}

By introducing the polar form (\ref{eq: scaled_wf_polar}) of the scaled wave function into the 
scaled equation (\ref{eq: Scaled Sch-Lan-1}) and then decomposing into imaginary and real parts, 
one easily obtains  (\ref{eq: tran_imag_part}) and (\ref{eq: tran_real_part}) but with $\ti{\hbar}$
instead of $ \sqrt{\ep} \hbar$.
Then, from Eq. (\ref{eq: tran_imag_part}), the continuity equation is readily obtained to be
\begin{eqnarray} \label{eq: bm1}
\frac{\pa \ti{\rho}}{\pa t} + \frac{\pa}{\pa x} ( \ti{\rho} v) &=& 0 , 
\end{eqnarray}
where
\begin{eqnarray}
\ti{\rho}(x, t) &=& R^2_{\ep}(x, t)
\end{eqnarray}
and
\begin{eqnarray} \label{eq: vel_field}
v(x, t) &=& \frac{1}{m} \frac{\pa S_{\ep}(x, t)}{\pa x}
\end{eqnarray}
are the probability density and the corresponding velocity field, respectively.
From Eq. (\ref{eq: tran_real_part}) and taking into consideration the quantum potential defined
by Eq. (\ref{qpe}),
%
%\begin{eqnarray}
%Q_{\ep}(x, t) &=& - \frac{\hbar^2}{2m} \frac{1}{R_{\ep}} \frac{\pa^2 R_{\ep}}{\pa x^2}
%\end{eqnarray}
%
one finds the corresponding Hamilton-Jacobi equation for the phase
\begin{eqnarray}
- \frac{\pa S_{\ep}}{\pa t} &=&  
 \frac{1}{2m}  \left( \frac{\partial  S_{\ep}}{\partial x} \right)^2
 + V(x, t) + V_r(x, t) + \gamma ( S_{\ep} - \langle S_{\ep} \rangle ) + \ti{Q}(x, t) ,
%\qquad \ti{Q} = \ep Q_{\ep} ,
\end{eqnarray}
with  $\ti{Q} = \ep Q_{\ep}$.  By taking the partial derivative with respect to the space 
coordinate and using (\ref{eq: vel_field}), the differential equation for the velocity field is given by
\begin{eqnarray} \label{eq: bm2}
\frac{dv}{dt}&=& \frac{\pa v}{\pa t} + v \frac{\pa v}{\pa x} = - \frac{1}{m} \frac{\pa}{\pa x} \left( V(x, t) + V_r(x, t) + \ti{Q}(x, t)  \right) - \gamma v  ,
\end{eqnarray}
which is the classical equation of motion but with the additional term $\ti{Q}$ responsible 
for non-classical effects. It is quite usual to solve Eqs. (\ref{eq: bm1}) and (\ref{eq: bm2}) 
by imposing a time-dependent Gaussian ansatz for the probability density 
\cite{NaMi-book-2017, ZaPlJD-AP-2015},
\begin{eqnarray} \label{eq: rho_ansatz}
\ti{\rho} (x, t) &=& \frac{1}{\sqrt{2\pi} ~ \ti{\sigma}(t)} \exp \left[ -\frac{(x-x_t)^2}{2 \ti{\sigma}^2(t)} \right] ,
\end{eqnarray}
where $x_t = \int dx ~ x \ti{\rho}(x, t) $ is the time dependent expectation value of the 
position operator which follows the center of the Gaussian wave packet and $\ti{\sigma}(t)$ gives 
its width. 
Eq. (\ref{eq: rho_ansatz}) satisfies  the continuity equation (\ref{eq: bm1}) for
\begin{eqnarray} \label{eq: Gauss-vel-field}
v(x, t) &=& \frac{\dot{\ti{\sigma}}}{\ti{\sigma}} (x-x_t) + \dot{x}_t ,
\end{eqnarray}
from which scaled trajectories are generated and expressed as 
\begin{eqnarray} \label{eq: scaled-trajsield}
x(x^{(0)}, t) &=& x_t + (x^{(0)} - x_0) \frac{\ti{\sigma}(t)}{\sigma_0} ,
\end{eqnarray}
with $x^{(0)}$ being the initial condition for the coordinate, $x_0$ the initial value for $x_t$ and $\sigma_0 = \ti{\sigma}(0)$. 
When $\epsilon = 1$, we have the quantum trajectories of Bohmian mechanics.
The structure of Eq. (\ref{eq: scaled-trajsield}) is typical in this dynamics where a scaled trajectory 
is formed by a classical trajectory $x_t$ plus a term involving the wave character
of the non-classical particle through its scaled wave packet width. This is known in the literature as
{\it dressing scheme} \cite{NaMi-book-2017}. Now, by replacing Eqs. (\ref{eq: rho_ansatz}) 
and (\ref{eq: Gauss-vel-field}) into Eq. (\ref{eq: bm2}), and then Taylor expanding the interaction
potential around $x_t$ up to second order and using the condition for linear independence 
of different powers of $x-x_t$, a classical Langevin equation for the position of the center 
of the wave packet and a second order differential equation in time for the width are easily derived  
\begin{eqnarray}
\ddot{x_t} + \gamma \dot{x_t} + \frac{1}{m} \left( F_r(t) + \frac{\pa V}{\pa x} \bigg|_{x=x_t} \right) &=& 0 , \label{eq: xbar}  \\
\ddot{ \ti{\sigma} } + \gamma \dot{ \ti{\sigma} } - \frac{\ti{\hbar}^2}{4 m^2 \ti{\sigma}^3} + 
\frac{ \ti{\sigma} }{m} \frac{\pa^2 V}{\pa x^2} \bigg|_{x=x_t} &=& 0 , \label{eq: delta} 
\end{eqnarray}
where a linear form (\ref{eq: random_pot}) is explicitely assumed for the random potential.
From these equations it is clear seen that the transition parameter $\ep$ and the friction 
coefficient affect the width of the wave packet whereas, as expected from Ehrenfest theorem, 
the  motion of its center is not altered by $\epsilon$ and, therefore, $x_t$ follows a 
classical trajectory. Notice that the scaled Planck's contant appears only in the differential
equation for the wave packet width implying that, with $\ep$, the quantum character of its time evolution 
is gradually lost. It is interesting to stress here that for potentials of at most quadratic order this dynamics 
is exact, meaning that the Gaussian ansatz is the exact solution of the transition wave equation for these cases.

If we neglect the random force term $F_r(t)$ and consider only dissipation and 
the second order interaction potential
\begin{eqnarray} \label{eq: quad_taylor}
V(x, t) & = & V_0(t) + V_1(t) x + \frac{1}{2} V_2(t) x^2 ,
\end{eqnarray}
then, from the previous two equations we have
\begin{eqnarray}
\ddot{x_t} &=& - \gamma \dot{x_t} - \frac{V_1(t)}{m} - \frac{V_2(t)}{m} x_t . 
\label{eq: xbar_quadratic} \\
\ddot{ \ti{\sigma} } &=& - \gamma \dot{ \ti{\sigma} } + \frac{ \ti{\hbar}^2}{4 m^2 \ti{\sigma}^3} - 
\frac{V_2(t)}{m} \ti{\sigma}  . \label{eq: delta_quadratic}  
\end{eqnarray}
%
%where $V_1(t)$ and $V_2(t)$ are the first and second Taylor coefficients of the interaction potential
%$V(x,t)$ evaluated at $x_t$, i.e.,

Provided that $V_1$ and $V_2$ are time-independent, then the solution of  
Eq. (\ref{eq: xbar_quadratic}) is analytical and given by
\begin{eqnarray} \label{eq: xbar_quadratic_sol}
x_t &=& - \frac{V_1}{V_2} + \left( x_0 + \frac{V_1}{V_2} \right) \left[ \cosh \Omega t + \frac{\gamma}{2} \frac{\sinh \Omega t}{\Omega} \right] e^{-\gamma t /2} + \dot{x}_0 ~ \frac{\sinh \Omega t}{\Omega}~e^{-\gamma t /2} ,
\end{eqnarray}
with
\begin{eqnarray} \label{eq: Omega}
\Omega &=& \sqrt{-V_2/m + \gamma^2/4} .
\end{eqnarray}
On the contrary, the solution of  Eq. (\ref{eq: delta_quadratic}) is not found in an analytical 
way. However, for the non-dissipative or frictionless case $ \gamma = 0 $, provided that $V_2$ 
is independent of time, its solution is given by
\begin{eqnarray} \label{eq: sigma_nondis}
\ti{\sigma}(t) &=& \sigma_0 \sqrt{ \cosh^2(\omega t) + \frac{\ti{\hbar}^2}{4 m^2 \omega^2 \sigma_0^4} \sinh^2(\omega t) } 
\end{eqnarray}
for $\dot{\ti{\sigma}}(0) =0 $ and where $\omega = \sqrt{-V_2/m}$.
For the classical regime, $ \ep = 0$, and $V_2$ independent of time, the  solution of 
Eq. (\ref{eq: delta_quadratic}) is given by
\begin{eqnarray} \label{eq: delta_quadratic_cl}
\sigma_{\cl}(t) &=& \sigma_0 \left( \cosh \Omega t + \frac{\gamma}{2} \frac{\sinh \Omega t}{\Omega} \right) e^{-\gamma t /2} + \dot{\sigma}_0 \frac{\sinh \Omega t}{\Omega} ~  
e^{-\gamma t /2}
\end{eqnarray}
which leads to 
\begin{eqnarray} \label{eq: delta_quadratic_cl_nondis}
\sigma_{\cl}(t) &=& \sigma_0 ~ \cosh(\omega t) +  \dot{\sigma}_0 ~ \frac{ \sinh(\omega t) }{\omega}
\end{eqnarray}
in the non-dissipative case. In these equations $ \dot{\sigma}_0 $ stands for the initial value of $\dot{\ti{\sigma}}(t)$. 

Now, from Eq. (\ref{eq: scaled-trajsield}), the  difference between two typical scaled trajectories
can be expressed as  
\begin{eqnarray} \label{eq: trajs_diff}
x(x_1^{(0)}, t) - x(x_2^{(0)}, t) &=& (x_1^{(0)} - x_2^{(0)}) 
\frac{\ti{\sigma}(t)}{\ti{\sigma_0}}~ .
\end{eqnarray}
Thus, trajectories diverge during the  time evolution revealing the non-crossing property of trajectories.
Notice that this property is even valid in the classical regime and will be used to provide a criterion 
for the tunneling process.
%Note: Both theory are equivalent for the non-dissipative case $\gamma = 0$.

%----------------------
\subsection{The CK approach}
%---------------------

By introducing again the polar form (\ref{eq: scaled_wf_polar}) of the scaled wave function into 
the scaled CK equation (\ref{eq: CK Scaled Sch}) and then splitting into imaginary and real parts, 
one obtains respectively
\begin{eqnarray}
\frac{\pa}{\pa t} R_{\ep}^2 + \frac{\pa}{\pa x} \left( R_{\ep}^2 \frac{1}{m} \frac{\pa S_{\ep}}{\pa x}  e ^{-\gamma t} \right) &=& 0,  \label{eq: CK_de1}
\\
\frac{\pa S_{\ep}}{\pa t} + \frac{1}{2m}  \left( \frac{\partial  S_{\ep}}{\partial x} \right)^2 e^{-\gamma t}
 + V(x, t) e^{\gamma t} + \ti{Q}(x, t) &=& 0 , \label{eq: CK_de2}
\end{eqnarray}
where 
\begin{eqnarray} \label{CK_scaled qp}
\ti{Q}(x, t) &=& - \frac{\ti{\hbar}^2}{2m} \frac{1}{ R_{\ep} } \frac{\pa^2 R_{\ep} }{\pa x^2 } e^{-\gamma t}
\end{eqnarray}
is the quantum potential in the CK framework. Furthermore, by introducing the scaled velocity field as 
\begin{eqnarray} \label{CK_vel field}
v(x, t) &=& \frac{1}{m} \frac{\pa S_{\ep}}{\pa x}  e ^{-\gamma t}   ,
\end{eqnarray}
Eq. (\ref{eq: CK_de1}) can be rewritten as the continuity equation (\ref{eq: bm1}). By taking 
the space partial derivative of Eq. (\ref{eq: CK_de2}) and using the velocity filed 
(\ref{CK_vel field}), one readily obtains
\begin{eqnarray} \label{CK_bm2}
\frac{\pa v}{\pa t} + v \frac{\pa v}{\pa x} &=& - \frac{1}{m} \frac{\pa}{\pa x} \left( V(x, t) + \ti{Q}(x, t) e ^{-\gamma t} \right) - \gamma v  .
\end{eqnarray}

If the Gaussian ansatz (\ref{eq: rho_ansatz}) is again assumed for the solution of  the continuity 
equation, the same velocity field (\ref{eq: Gauss-vel-field}) and scaled trajectories 
(\ref{eq: scaled-trajsield}) are reached.  Furthermore, one can rewrite Eq. (\ref{CK_bm2}) as
\begin{eqnarray} \label{CK_bm2_2}
\bigg[ \ddot{ \ti{\sigma} } + \gamma \dot{ \ti{\sigma} } -\frac{\ti{\hbar}^2}{4 m^2 \ti{\sigma}^3} e^{-2\gamma t} \bigg]
(x-x_t) + \ti{\sigma} ( \ddot{x}_t + \gamma \dot{x}_t )
&=& -\frac{ \ti{\sigma} }{m} \frac{\pa V}{\pa x} .
\end{eqnarray}
Following the same procedure as before, the corresponding differential equations for the 
center of the wave packet and width are now given by
%Now, by Taylor expanding the interaction potential around $x_t$ up to the second order, 
%Eq. (\ref{eq: quad_taylor}), and using the condition for linear independence of different powers 
%of $x-x_t$, an equation for the center of the Gaussian packet and a second order differential
%equation in time for the width are easily derived  
%
\begin{eqnarray}
\ddot{x_t} &=& - \gamma \dot{x_t} - \frac{V_1(t)}{m} - \frac{V_2(t)}{m} x_t , 
\label{eq: CK_xbar_quadratic} \\
\ddot{ \ti{\sigma} } &=& - \gamma \dot{ \ti{\sigma} } + \frac{ \ti{\hbar}^2}{4 m^2 \ti{\sigma}^3} e^{-2\gamma t} - \frac{V_2(t)}{m} \ti{\sigma}  . \label{eq: CK_delta_quadratic}  
\end{eqnarray}
for the quadratic potential (\ref{eq: quad_taylor}).
These equations are quite similar to those previously reached except the time exponential factor.

From (\ref{eq: xbar_quadratic}) and (\ref{eq: CK_xbar_quadratic}), one sees that the 
differential equation for the motion of the center of the wave packet is the same in both
approaches. However, the differential equations for the width differ again by a time exponential
factor. After Ehrenfest' theorem, it is not surprising to observe that the nonlinearity displayed by 
the Kostin approach is not manifested in a trajectory description of the quantum dynamics. 
%It is clear from Eqs. (\ref{eq: delta_quadratic}) and 
%(\ref{eq: CK_delta_quadratic}) that for the classical case, $\epsilon=0$, both equations are the 
%same meaning that both models are classically identical.  

With the initial conditions $\ti{\sigma}(0) = \sigma_0$ and $\dot{\ti{\sigma}}(0) = 0$, 
and $V_2(t)$ independent on time, the solution of  Eq. (\ref{eq: CK_delta_quadratic}) is 
analytical and given by
\begin{eqnarray} \label{eq: CK_width_quandratic}
\ti{\sigma}(t) &=& \sigma_0~e^{-\gamma t /2}~ 
\sqrt{ \left( \cosh \Omega t + \frac{\gamma}{2} \frac{\sinh \Omega t}{\Omega} \right)^2 + \frac{\ti{\hbar}^2}{4m^2 \sigma_0^4} \frac{\sinh^2 \Omega t}{\Omega^2} } ,
\end{eqnarray}
$\Omega$ being defined by Eq. (\ref{eq: Omega}). 
In the classical limit, Eq. (\ref{eq: CK_width_quandratic}) reduces to
\begin{eqnarray} \label{eq: cl_config_width}
\sigma_{\cl}(t) &=& \left( \cosh \Omega t + \frac{\gamma}{2} \frac{\sinh \Omega t}{\Omega} \right) \sigma_0 ~  e^{-\gamma t /2} 
\end{eqnarray}
and to Eq. (\ref{eq: sigma_nondis}) in the frictionless case.
% a not an unexpected result noting the equivalence of both models for the non-dissipation case. 

%=====================================================
%=====================================================

\subsection{Distribution function for actual momentum}

In a trajectory description of quantum mechanics, the system is govened by 
its wave function and position. 
%Particle trajectories $x(x^{(0)}, t)$ is specified by the guidance equation.
%
Assuming that the initial distribution function for particle positions is given by the Born rule, 
it is concluded by means of the continuity equation that the Born rule holds at any time,
\begin{eqnarray*}
\rho(x, t) &=& \int dx^{(0)} \rho(x^{(0)}, 0) ~ \delta \left( x - x(x^{(0)}, t) \right) ~.
\end{eqnarray*}

The probability distribution function for a particle property $f$ is given by \cite{Le-book-2002}
\begin{eqnarray} \label{eq: dis_func}
\Pi(f) &=& \int dx^{(0)} ~\rho(x^{(0)}, 0) ~ \delta \left( f - f(x^{(0)}) \right)~, 
\end{eqnarray}
where $ f(x^{(0)}) $ is the value of $f$ along the trajectory $ x(x^{(0)}, t) $.
Since the Bohmian momentum filed is given by $ p(x, t) = m ~ \dot{x}(x, t) $, from
Eq. (\ref{eq: dis_func}) one has
\begin{eqnarray} \label{eq: mom_dis}
\Pi(p, t) &=& \int dx^{(0)} ~\rho(x^{(0)}, 0) ~ \delta \left( p -  m ~ \dot{x}(x, t) \bigg|_{ x = x(x^{(0)}, t) } \right)
\end{eqnarray}
for the {\it actual} momentum \cite{Holland-book-1993} distribution function. 
 Thus, by using Eqs. (\ref{eq: rho_ansatz}), (\ref{eq: Gauss-vel-field}) and 
 (\ref{eq: scaled-trajsield}) into (\ref{eq: mom_dis}) and the definition $p_t = m \dot{x}_t$, 
 we have that
\begin{eqnarray} 
\ti{\Pi}(p, t) &=& \frac{1}{ \sqrt{2\pi \sigma_0^2} } \int dx^{(0)} ~ \exp \left[ -\frac{(x^{(0)}-x_0)^2}{2\sigma_0^2} \right]
~ \delta \left( p - p_t - m \frac{\dot{ \tilde{\sigma}}(t)}{\sigma_0} (x^{(0)}-x_0)  \right) 
\nonumber \\
&=&  
\frac{1}{ \sqrt{2\pi \sigma_0^2} } \int dy ~ \exp \left[ -\frac{y^2}{2\sigma_0^2} \right]
~ \delta \left( m \frac{\dot{ \tilde{\sigma}}(t)}{\sigma_0} y + p_t - p )  \right)  
\nonumber
\\
&=&
\frac{1}{ \sqrt{2\pi \tilde{\Sigma}(t)^2} } ~ \exp \left[ -\frac{(p-p_t)^2}{2\tilde{\Sigma}(t)^2} \right] , 
\label{eq: mom_dis__generalform} 
\end{eqnarray}
where
\begin{eqnarray} \label{eq: sigmaP_general}
\tilde{\Sigma}(t) &=& m \dot{\tilde{\sigma}}(t)
\end{eqnarray}
is the width of the momentum distribution. 
This equation shows that the actual momentum has a Gaussian shape around the momentum 
$p_t$ with width $\tilde{\Sigma}(t)$. For the classical regime $\ep = 0$, 
the width is given by Eq. (\ref{eq: cl_config_width}) and thus from Eq. (\ref{eq: sigmaP_general}),
one has the following width
\begin{eqnarray} \label{eq: sigmaP_cl}
\tilde{\Sigma}_{\text{cl}}(t) &=& m \sigma_0 \omega^2 \frac{ \sinh \Omega t }{ \Omega } 
e^{-\gamma t/2}~,
\end{eqnarray}
for the width of the actual momentum in the classical regime. Furthermore, since
$ \bar{\Sigma}_{\text{cl}}(0)=0 $, initially all particles in the classical ensemble have 
the same momentum $p_0$. From this, we will give a good criterion for tunnelling.

%=====================================================
%=====================================================

\section{Tunnelling from a parabolic repeller potential in the presence of an oscillatory electric field}
Let us consider now the dissipative tunnelling dynamics of charged particles with charge 
$q$ described  by a Gaussian wave packet (\ref{eq: rho_ansatz}) from a parabolic repeller or 
inverted harmonic oscillator potential and under the action of an oscillatory electric field
\begin{eqnarray} \label{eq: pot}
V(x, t) &=& q E_0 \cos( \omega_0 t + \phi) ~ x - \frac{1}{2}  m \omega^2 x^2  .
\end{eqnarray}
where $\omega$ is the frequency of the oscillator, $m$ is the mass of the harmonic
oscillator and $E_0$, $\omega_0$ and $\phi$ give the amplitude, frequency and phase of the 
applied field, respectively.
For this potential, the equation of motion for the center of the Gaussian wave packet, 
given by Eq. (\ref{eq: xbar}), reads now  as 
\begin{eqnarray} \label{eq: center_PROS}
\ddot{x_t} + \gamma \dot{x_t} - \omega^2 x_t &=& - \frac{q E_0}{m}\cos( \omega_0 t + \phi), 
\end{eqnarray}
while Eqs. (\ref{eq: CK_delta_quadratic}) and (\ref{eq: delta_quadratic})  
appearing in both approaches transform to
\begin{eqnarray} 
\ddot{ \ti{\sigma} }  + \gamma \dot{ \ti{\sigma} } - \frac{ \ti{\hbar}^2}{4 m^2 \ti{\sigma}^3} e^{-2\gamma t} - \omega^2 \ti{\sigma} = 0, 
\label{eq: CK_width_PROS} \\
\ddot{ \ti{\sigma} }  + \gamma \dot{ \ti{\sigma} } - \frac{ \ti{\hbar}^2}{4 m^2 \ti{\sigma}^3} - \omega^2 \ti{\sigma} = 0
\label{eq: Kostin_width_PROS}
\end{eqnarray}
which clearly show that the width of the Gaussian wave packet does not depend on the applied 
field parameters. The solution of Eq. (\ref{eq: center_PROS}) is given by 
\begin{eqnarray} \label{eq: xbar_quadratic_sol}
x_t &=& \left[ x_0 \left( \cosh \Omega t + \frac{\gamma}{2} \frac{\sinh \Omega t}{\Omega}
\right) + \dot{x}_0 ~ \frac{\sinh \Omega t}{\Omega} \right] e^{-\gamma t /2}
\nonumber \\
&+& e^{-\gamma t/2} \frac{q E_0 /m }{ \gamma^2 \omega_0^2 + ( \omega_0^2 + \omega^2  )^2 }
\left[
\left( \frac{\gamma^2}{2} + \omega_0^2 + \omega^2 \right) \frac{\sinh \Omega t}{\Omega} + \gamma \cosh \Omega t
\right] \omega_0 \sin \phi
\nonumber \\
&+& e^{-\gamma t/2} \frac{q E_0 /m }{ \gamma^2 \omega_0^2 + ( \omega_0^2 + \omega^2  )^2 }
\left[
 ( \omega_0^2 - \omega^2 ) \frac{\gamma}{2} \frac{\sinh \Omega t}{\Omega} - ( \omega_0^2 + \omega^2 ) \cosh \Omega t
\right] \cos \phi
\nonumber \\
&+&
\frac{q E_0/m }{ \gamma^2 \omega_0^2 + ( \omega_0^2 + \omega^2  )^2 }
\left[
(\omega_0^2 + \omega^2) \cos (\omega_0 t + \phi)
- \gamma \omega_0 \sin (\omega_0 t + \phi)
\right],
\end{eqnarray}
where the last line is the particular solution of Eq. (\ref{eq: center_PROS}) with $\Omega$ given by
Eq. (\ref{eq: Omega}).
%
%\begin{eqnarray} \label{eq: Omega}
%\Omega &=& \sqrt{\omega^2 + \frac{\gamma^2}{4}} .
%\end{eqnarray}
%

The time-dependent transmission probability for incidence from left to right 
(from negative to positive values of the coordinate) of the parabolic barrier 
is well known to be \cite{BaJa-JPA-1992,Pa-JPA-1990,Pa-JPA-1997}
\begin{eqnarray} \label{eq: tran_prob}
T(t) &=& \frac{ B(t) }{ \int_{-\infty}^{x_m} dx~\rho(x, 0) }
\end{eqnarray}
with
\begin{eqnarray} \label{eq: B_coeff}
B(t) &=& \int_0^t dt' ~ j(x_d, t') = \int_{x_d}^{\infty} dx~[\rho(x, t) - \rho(x, 0)]
\end{eqnarray}
and $x_m$ is the location of the barrier maximum, or top $x_m  = 0$. Note that $x_d$ can be 
any point on the right side of the barrier. In the stationary regime, where the transmission probability 
becomes constant, its value is independent of the choice of $x_d$. The second equality in 
Eq. (\ref{eq: B_coeff}) results from integrating the continuity equation.
For $x_d = x_m = 0$, Eq. (\ref{eq: tran_prob}) reduces to 
\begin{eqnarray} \label{eq: tran_prob_gauss}
T(t) &=& \frac{ {\text{erf}}( x_t / \sqrt{2} \ti{\sigma}(t) ) - {\text{erf}}( x_0 / \sqrt{2} \sigma_0 ) }{ {\text{erfc}}( x_0 / \sqrt{2} \sigma_0 ) }
\end{eqnarray}
for the wave packet (\ref{eq: rho_ansatz}). The transmission probability thus depends on the 
field parameters through $x_t$ in an indirect way and is written in terms of the error (erf) 
and its complementary (erfc) functions \cite{Gradshteyn-1980}.

%=====================================================
%=====================================================

\section{Results and discussion}

%A usual criterion to deal with tunnelling is the following: particle’s expectation energy must be 
%smaller than the barrier height. This criterion is used properly for conservative systems. However, in 
%non-conservative systems, including dissipative media, particles loss energy along time and 
%therefore this criterion should be modified. Authors of \cite{BaJa-JPA-1992} have used 
%this unreliable condition to study tunnelling by a parabolic repeller. Papadopolus 
%\cite{Pa-JPA-1997}, in the study of microwave-assisted tunnelling in a viscous medium, 
%introduced another criterion: classical particles starting from the center of the initial 
%packet and moving with kick momentum $p_0$ never pass to the other of the barrier. 
%In our formalism, there are however some classical particles with initial position $x^{(0)}>x_0$ 
%and the same momentum $p_0$ that can pass the height of the barrier to the other side. 

Our numerical calculations are carried out  in a system of units where $m=1$, $\hbar=1$ and 
$q=-1$. Furthermore, the parameters of the initial Gaussian wave packet 
and frequency of  the parabolic repeller are chosen to be $\sigma_0=1$, $p_0 = 1$, $x_0 = -10$ and 
$\omega = 0.2$, respectively. The friction coefficient and parameters describing the oscillatory
electric field are varied to study their mutual interference in this dissipative tunnelling dynamics.

%Figure 
\begin{figure} 
\centering
\includegraphics[width=8cm,angle=-90]{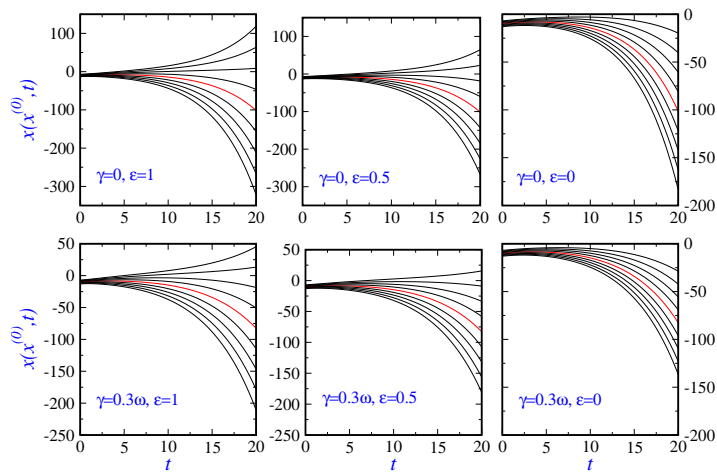}
\caption{(Color online) Scaled trajectories issued from a propagating Gaussian wave packet 
under the potentail $V(x, t) = qE_0 [\cos(\omega_0 t + \phi)] x - \omega^2 x^2 / 2$ 
in the CK approach. The red curve corresponds to the classical motion of the center 
of the wave packet. In each row, the friction coefficient $\gamma$ is 
constant: $\gamma = 0$  (first row) and $\gamma = 0.3 \omega$ (second row).
In each column, the dynamical regime given by $\epsilon$ is constant:  quantum regime, 
$\epsilon = 1$ (first column);  intermediate regime, $\epsilon = 0.5$ (second column); and 
classical regime, $\epsilon = 0$ (third column). Parameters have been fixed as follows: $\phi =0$, 
$\omega_0 = \omega = 0.2$, $E_0 = 0.1$, $x_0=-10$, $\sigma_0 = 1$ and $p_0 = 1$.
}
\label{fig: CK_trajs}
\end{figure}

In Figure \ref{fig: CK_trajs}, scaled trajectories are plotted in the CK framework 
for different dynamical regimes ruled by  $\epsilon$ for the non-dissipative, $\gamma=0$
(top panels), and dissipative, $\gamma=0.3\omega$ (bottom panels), cases. The field 
parameters are $E_0 = 0.1$, $\omega_0 = \omega = 0.2$ and $\phi =0$. %The initial parameters
%of the wave packet are $x_0=-10$, $\sigma_0 = 1$ and $p_0 = 1$.
A uniform distribution of initial positions in the range $[x_0-3\sigma_0, x_0+3\sigma_0]$
for the initial Gaussian probability density function $\ti{\rho}(x, 0)$ is used. In each plot, 
the red curve corresponds to the motion of center of the wave packet which is a classical 
path independent on $\epsilon$. Three different regimes are then identified: classical regime,
$\epsilon =0$, intermediate or transition regime, $\epsilon = 0.5$, and quantum regime with 
$\epsilon = 1$. 
%
%Figure
\begin{figure}
\centering
\includegraphics[width=8cm,angle=-90]{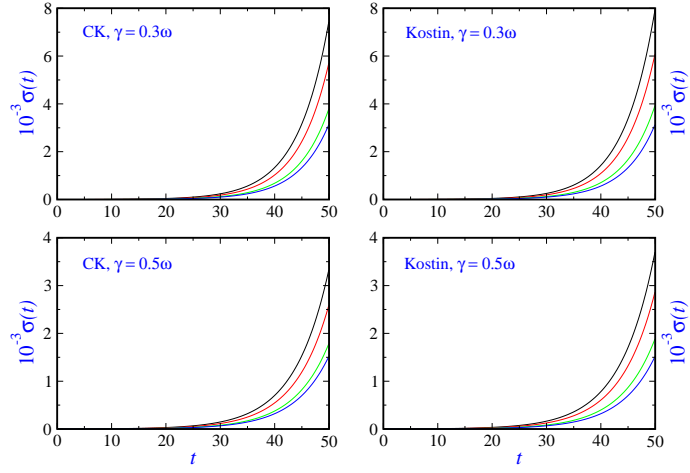}
\caption{
Width of the wave packet versus time in the CK (first column) and Kostin (second column) 
approaches for different values of the dynamical regime: $\epsilon = 1$ (black), 
$\epsilon = 0.5$ (red), $\epsilon = 0.1$ (green) and $\epsilon = 0$ (blue) and two values of 
friction: $\gamma = 0.3 \omega$ (first row) and $\gamma = 0.5 \omega$ (second row).
Values of the field parameters are the same as in Figure \ref{fig: CK_trajs}.}
\label{fig: width_time}
\end{figure}
As expected, classical trajectories (for $\epsilon = 0$) do not cross the barrier 
revealing the characteristics of tunnelling process.
%since no trajectory displays positive values with time.
However, in the transition to quantum regime, some particles pass the barrier. 
%positive and negative values of $x_t$ are clearly observed in the remaining 
%panels. 
In particular, some of them above the red curve which their initial positions are located 
in the right tail of the initial Gaussian wave packet. These plots clearly reveal that the 
tunnelling process is present. 
%
%Although not observed in this figure, dissipative trajectories also display a localization effect. 
Scaled trajectories coming from the Kostin approach have the same behavior.

The  time evolution of the wave packet width $\ti{\sigma}(t)$ is shown in Fig.\ref{fig: width_time}
for different values of the dynamical regime: $\epsilon = 1$ (black curve),  
$\epsilon = 0.5$ (red curve),  $\epsilon = 0.1$ (green curve) and 
$\epsilon = 0$ (blue curve) in the CK approach (first column) and the Kostin
approach (second column). Two values of dissipation $\gamma = 0.3 \omega$ (first row) 
and $\gamma = 0.5 \omega$ (second row) are chosen. In both cases, the widths increase
with time but always we have that $ \sigma_{\text{CK}}(t) < \sigma_{\text{Kostin}}(t) $ 
indicating that tunnelling is more important in the Kostin framework.
When passing from the quantum to classical regime, the corresponding widths decrease smoothly
leading to a reduction of the weight of the wave part of the scaled trajectories in the dressing
scheme and, therefore, to the increase of the decoherence.  This fact is reinforced 
with dissipation since the spread of the probability density diminishes showing a tendency to observe
localization.

\begin{figure}
\centering
\includegraphics[width=8cm,angle=0]{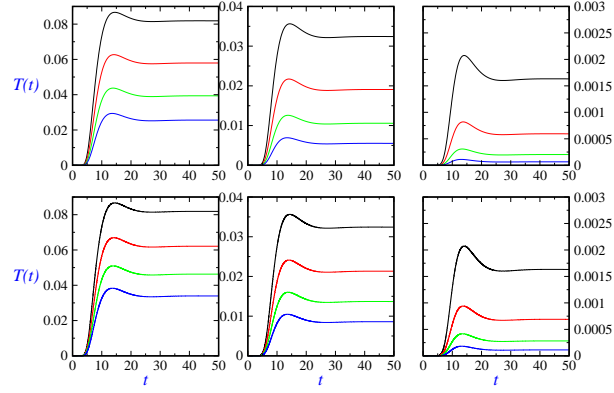}
\caption{(Color online)
Transmission probability as a function of time, $T(t)$, for different values of dissipation: 
$\gamma = 0$ (black curve), $\gamma = 0.1\omega$ (red curve), $\gamma = 0.2\omega$ 
(green curve) and $\gamma = 0.3\omega$ (blue curve). This probability is also plotted for 
$\epsilon = 1$ (first column), $\epsilon = 0.5$ (second column) and $\epsilon = 0.1$ (third column)   
in the  CK (first row) and Kostin (second row) approaches. Values of the field parameters are the 
same than in figure \ref{fig: CK_trajs}.}
\label{fig: tranprob_time}
\end{figure}

\begin{figure}
\centering
\includegraphics[width=8cm,angle=-90]{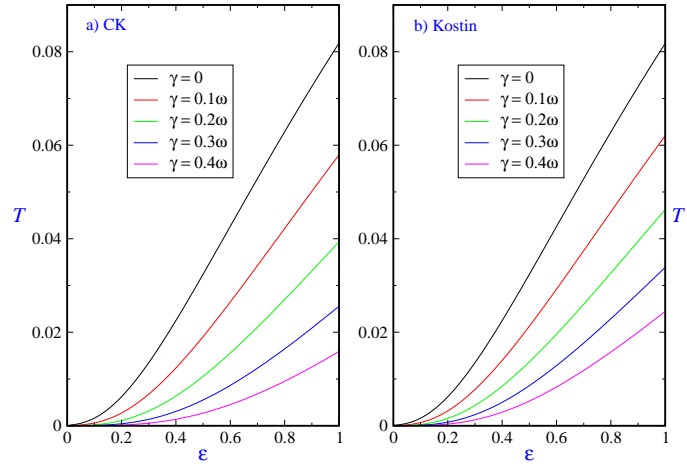}
\caption{(Color online) 
Transmission probability versus transition parameter $\epsilon$ for different values of 
$\gamma$ within a) the CK and b) Kostin framework. Values of the field parameters are 
the same as in Figure \ref{fig: CK_trajs}.}
 \label{fig: tranprob_ep}
\end{figure}

\begin{figure}
\centering
\includegraphics[width=8cm,angle=-90]{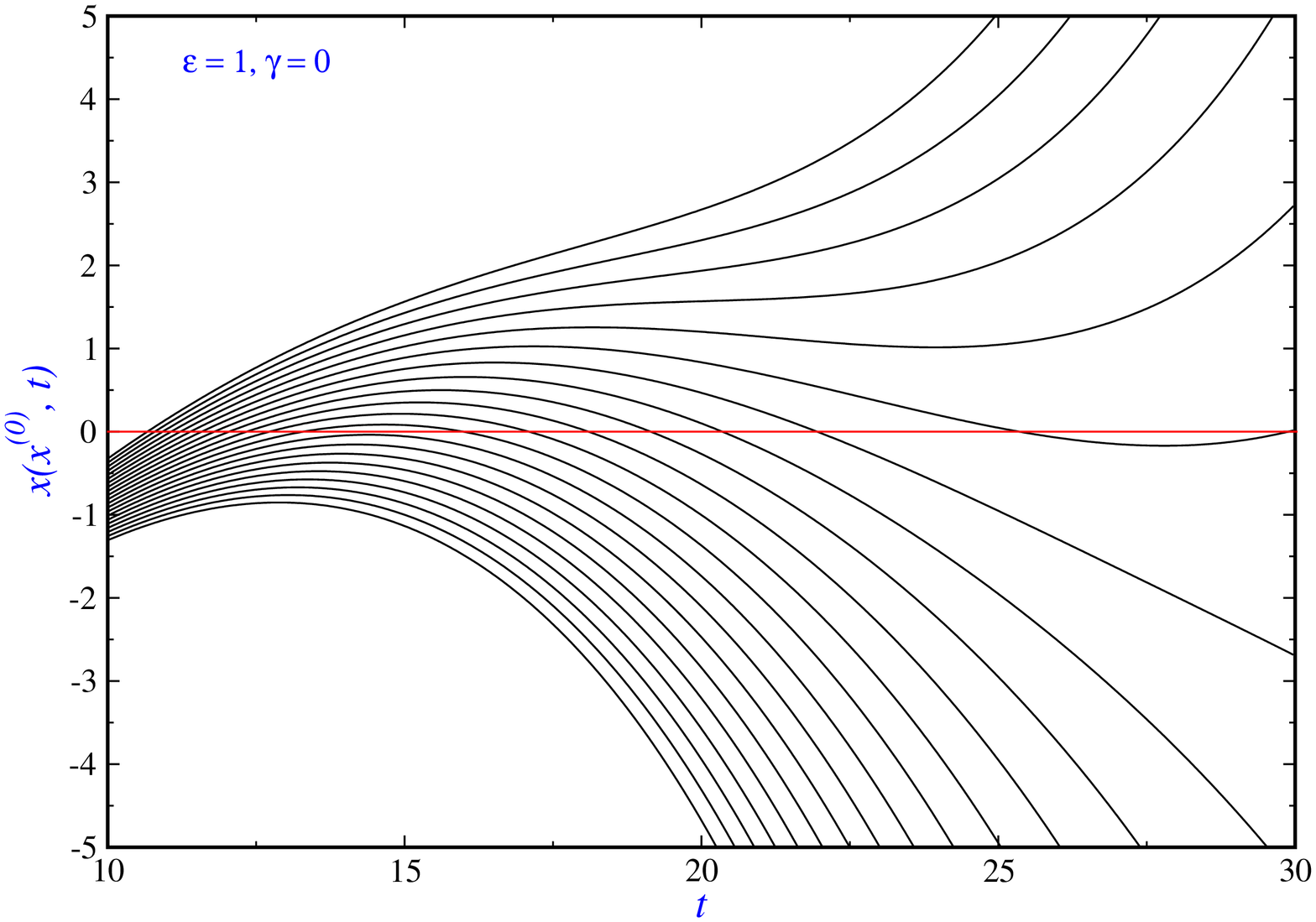}
\caption{(Color online)
A selection of scaled, $\ep=1$ (quantum), trajectories for the non-dissipative motion. 
The top of the parabolic barrier is shown by the red line. Values of the field parameters 
are the same as in figure \ref{fig: CK_trajs}.}
\label{fig: TURNED_TRAJS}
\end{figure}

Transmission probability is displayed versus time in Figure \ref{fig: tranprob_time}. 
Four different values of dissipation are chosen: $\gamma = 0$ (black curve), 
$\gamma = 0.1\omega$ (red curve), $\gamma = 0.2\omega$ (green curve) and 
$\gamma = 0.3\omega$ (blue curve). This probability is also plotted for different dynamical
regimes $\epsilon = 1$ (first column), $\epsilon = 0.5$ (second column) and $\epsilon = 0.1$ 
(third column)  in the  CK (first row) and Kostin (second row) frameworks. 
Several interesting features
are noticed. First, with friction, the transmission probabilities strongly decrease in both cases.
Second, with the transition parameter $\epsilon$, this probability also  decreases when passing 
from the quantum regime to a nearly classical regime ($\epsilon = 0.1$). And, finally, the Kostin
approach always displays a higher tunnelling process than the CK one. According to Eq. 
(\ref{eq: tran_prob_gauss}), the transmission probability is determined by 
${\text{erf}}( x_t / \sqrt{2} \ti{\sigma}(t) )$. Since the error function is an increasing function 
of its argument and $x_t$ is negative for tunnelling (see Fig. \ref{fig: CK_trajs}), 
then from $ \ti{\sigma}_{\text{CK}}(t) < \ti{\sigma}_{\text{Kostin}}(t) $, one sees 
that this probability is higher in the Kostin approach. This is also well illustrated in Fig. 
\ref{fig: tranprob_ep} where transmission probabilities are plotted as a function of the 
transition parameter at different dissipative values for the two cases. These 
probabilities are evaluated at $t=150$ where a constant value of $T$ is already reached.
As expected, these probabilities decrease with $\epsilon$ and $\gamma$ where the decoherence
process is playing a major role.
Moreover, for our choice of parameters, and according to Fig. \ref{fig: tranprob_time}
the transmission probability becomes constant after $t \approx 30$ and displays a maximum at 
very short times.  This maximum corresponds to trajectories passing through the barrier but 
after a while turn around. A selection of such trajectories are shown in Figure 
\ref{fig: TURNED_TRAJS} for the non-dissipative motion in the quantum regime.
%As expected dissipation decreases tunnelling. $T(t)$ decreases in transition from quantum to classical regime.
%
%Figure

%in a time $t_1$ which it has a constant value $T$ in such a time. Note that $t_1$ need not be 
%$t_0$ , but it can have any value greater than $t_0$. We choose instant $t_1 = 150$ for 
%computation of transmission probability. 
%We have shown transmission probability versus $\epsilon$ for different values of $\gamma$ in. 
\begin{figure}
\centering
\includegraphics[width=8cm,angle=0]{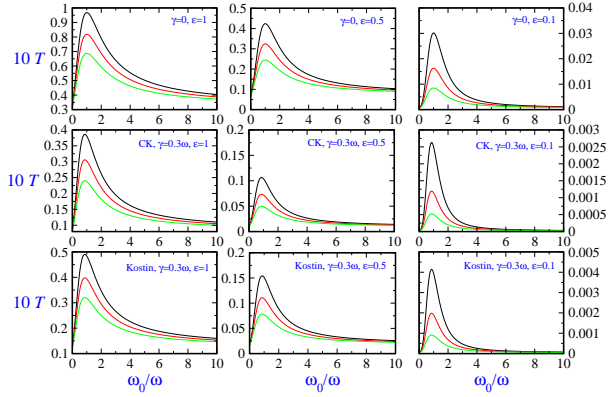}
\caption{(Color online) 
Transmission probability versus the frequency (in units of $\omega$) of the applied field 
for different amplitudes: $E_0 = 0.12$ (black curve), $E_0 = 0.1$ (red curve) and 
$E_0 = 0.08$ (green curve). The remianing parameters are the same as in figure \ref{fig: CK_trajs} 
but with $\phi=-\pi/2$. For easy comparison, in the first, second and third rows, the nondissipative
motion, the CK  and Kostin approach (for $\gamma = 0.3 \omega$) are considered, repectively.
The different dynamical regimes are shown in the three columns with $\epsilon = 1$, 
$\epsilon =0.5$ and $\epsilon =0.1$.}
\label{fig: tranprob_om0}
\end{figure}

\begin{figure}
\centering
\includegraphics[width=8cm,angle=-90]{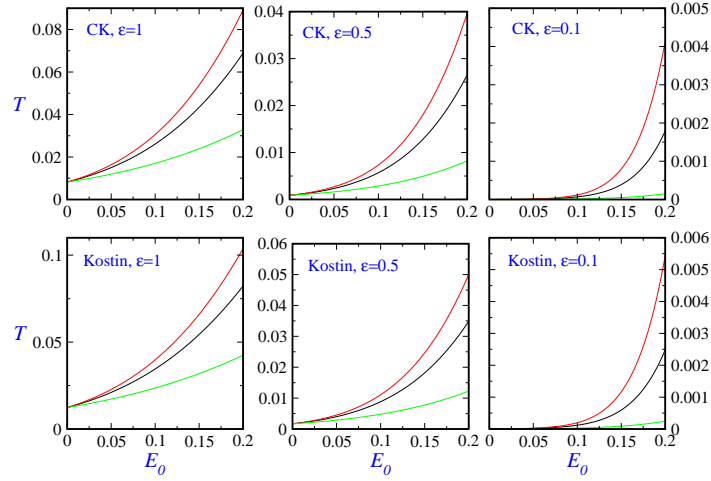}
\caption{(Color online)
Transmission probability versus the amplitude of the applied field for different values  
of the frequency: $\omega_0 = 0.5 \omega$ (black curve), $\omega_0 = \omega_{0,\text{res}}
\approx 0.86 \omega$ (red curve) and $\omega_0 = 3 \omega$ (green curve) 
at different dynamical regimes in the two approaches with $\gamma = 0.3 \omega$. 
The remaining parameters are the same as figure \ref{fig: CK_trajs} apart from $\phi=-\pi/2$.}
\label{fig: tranprob_E0}
\end{figure}

\begin{figure}
\centering
\includegraphics[width=8cm,angle=0]{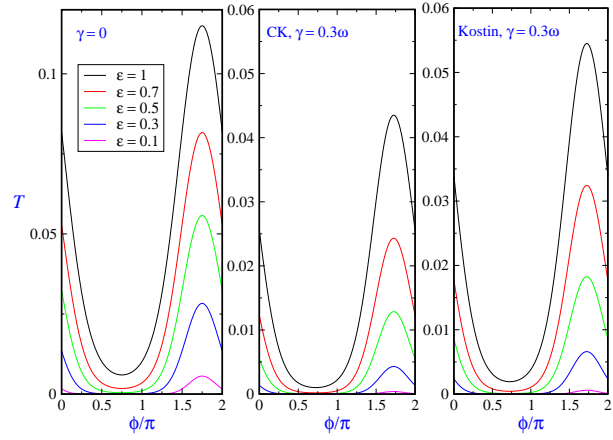}
\caption{(Color online)
Transmission probability versus the phase of the applied filed, $\phi$, for different dynamical 
regimes $\ep = 1$(black curve), $\ep = 0.7$(red curve), $\ep = 0.5$(green curve), 
$\ep = 0.3$(blue curve) and $\ep = 0.1$(magenta curve) with $\gamma = 0.3 \omega$. Other parameters are the same as those 
of figure \ref{fig: CK_trajs}.}
\label{fig: tranprob_phi}
\end{figure}

Let us analyze now  the effect of the oscillatory field in this dissipative tunnelling dynamics.
As has been shown before, the width $\ti{\sigma}(t)$ of the wave packet does not depend on the 
field parameters. Therefore, the variation of the tunnelling probability with the field 
parameters comes solely through $x_t$. Figures \ref{fig: tranprob_om0}, \ref{fig: tranprob_E0} and 
\ref{fig: tranprob_phi} display the dependence of transmission probability to the 
field parameters $\omega_0$, $E_0$ and $\phi$. In all of these figures, the transmission 
probability is always higher in the Kostin approach at the same friction parameter. 
Figure \ref{fig: tranprob_om0} shows the transmission probabilities versus the frequency 
(in units of $\omega$) of the applied field for different amplitudes: $E_0 = 0.12$ (black curve), 
$E_0 = 0.1$ (red curve) and $E_0 = 0.08$ (green curve) with $\phi=-\pi/2$.  
With this initial phase, the field is a pure sine function of $(\omega_0 t)$. For easy comparison, 
in the first, second and third rows, the nondissipative motion, the CK and Kostin approach
(for $\gamma = 0.3 \omega$) are considered, repectively. 
The different dynamical regimes are also shown in the three columns 
with $\epsilon = 1$, $\epsilon =0.5$ and $\epsilon =0.1$. The initial values of the Gaussian wave
packet are the same as in Fig. \ref{fig: CK_trajs}. A gradual decreasing of tunnelling is seen with
$\epsilon$, that is, when approaching the classical regime. In the absence of the field which this
occurs when $\omega_0 = 0$, the transmission probability is different from zero in the nonclassical
regime. A maximum is again observed in all cases but this time is not due to the back recrossing
of the scaled trajectories. This maximum is now attributed to a resonant transmission. As expected,
in the absence of friction, the  resonance takes place when $\omega_0 = \omega$. 
On the contrary, for a viscous medium, this  resonant mechanism is observed for 
$\omega_0 < \omega$. In our case, $ \omega_{0,\text{res}} \approx 0.86 \omega $.
Furthermore, the role of the field amplitude is just the reverse of  the friction, when increasing 
its value, the corresponding probabilities also increase in a nonlinear way. 
In the near classical regime $\epsilon = 0.1$, the tunnelling is very small. 
%but different from zero for $E_0 > 0.1$. 
All of these features are better illustrated in Fig. \ref{fig: tranprob_E0} 
where transmission probabilities are plotted versus the amplitude of the applied field for 
different values  of the frequency: $\omega_0 = 0.5 \omega$ (black curve), 
$\omega_0 = \omega_{0,\text{res}}\approx 0.86 \omega$ (red curve) and 
$\omega_0 = 3 \omega$ (green curve) at different dynamical
regimes in the two approaches with $\gamma = 0.3 \omega$. The resonance or red curve gives the 
maximum value of the transmission probability.

The behaviour of the tunnelling process as a function of the initial phase of the applied field
is plotted in Fig. \ref{fig: tranprob_phi}. Transmission probabilities versus $\phi$ for 
different dynamical regimes $\ep = 1$(black curve), $\ep = 0.7$(red curve), $\ep = 0.5$(green curve), 
$\ep = 0.3$(blue curve) and $\ep = 0.1$(magenta curve) for the frictionless motion, and the
CK and Kostin approaches with $\gamma = 0.3 \omega$.
In this figure, a resonant behaviour is again seen for all the dynamical regimes. The maximum
corresponds to $\phi \approx 1.75 \pi$ with $\omega_0 = 0.2$.

\begin{figure}
\centering
\includegraphics[width=8cm,angle=-90]{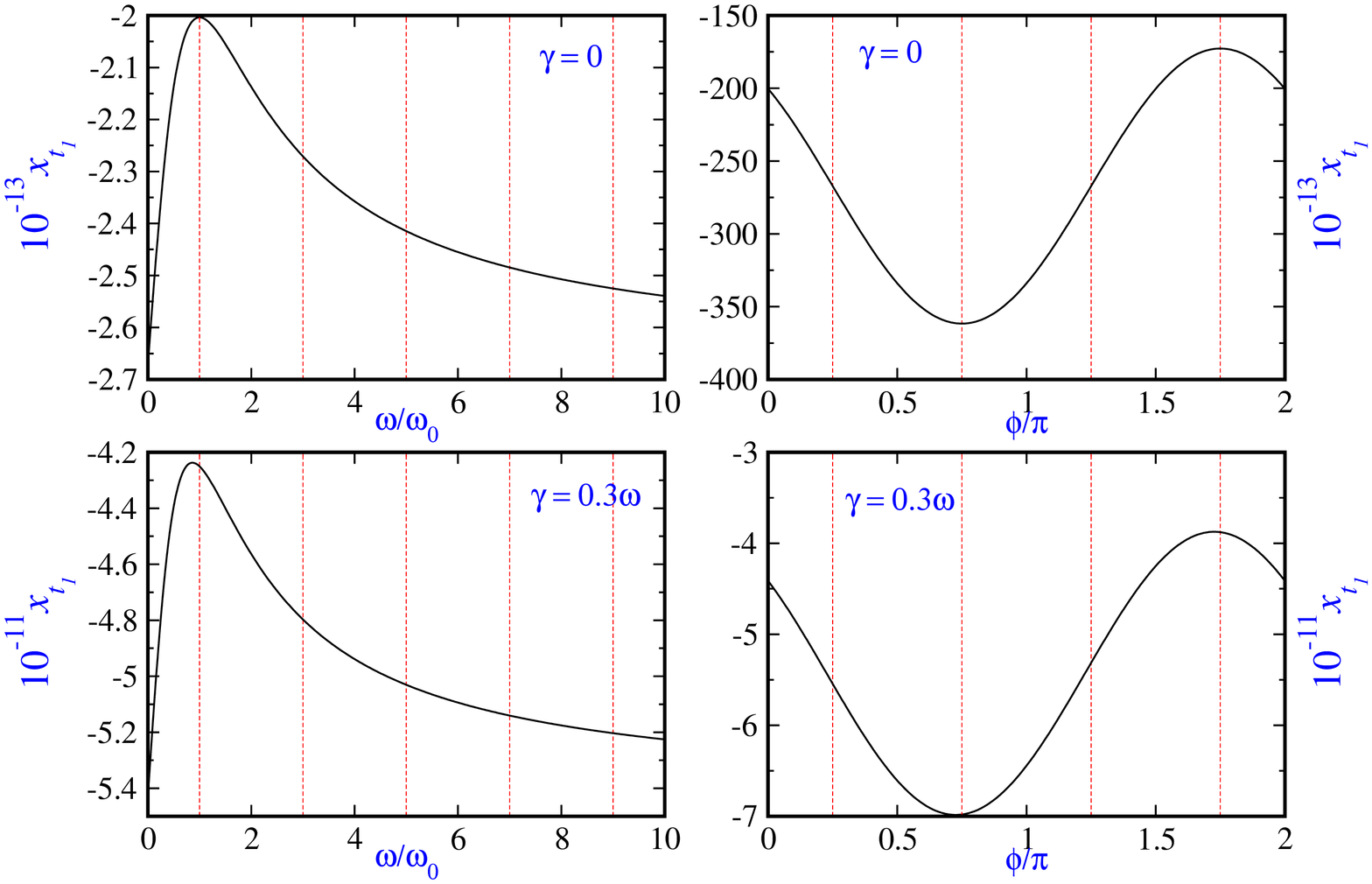}
\caption{(Color online)
Position of the wave packet center at time $t_1 = 150$ in which the transmission probability is 
computed versus the frequency $\omega_0$ (left column) and  phase $\phi$ (right column) field. 
The remaining parameters are the same as those of figure \ref{fig: CK_trajs}. 
Grid lines are included for better inspection.
In the frictionless motion, the maxima of $x_{t_1}$ versus $\omega_0$  and $\phi$ are located at 
$\omega_0 = \omega$ and $\phi = 1.75 \pi$, respectively. For $\gamma = 0.3 \omega$, the 
maxima are located at values $\omega_0 = 0.86 \omega$ and  $\phi \approx 1.73 \pi$, respectively. 
$x_{t_1}$ becomes minimum for $\phi = 0.75 \pi$.}
\label{fig: xt_om0phi}
\end{figure}

As commented above, the transmission probability depends on the field parameters only through
the classical trajectory $x_t$. In order to understand the location of the resonance of this probability
versus $\omega_0$ and $\phi$, a close inspection to $x_t$ should be carried out. To this end,
in Fig. \ref{fig: xt_om0phi},  the center of the wave packet is plotted at  $t_1 = 150$ versus 
$\omega_0$ and $\phi$ separately. In the frictionless motion, the maxima of $x_{t_1}$ versus $\omega_0$  
and $\phi$ are located at $\omega_0 = \omega$ and $\phi = 1.75 \pi$, respectively. On the contrary,
for $\gamma = 0.3 \omega$, the maxima are located at values $\omega_0 = 0.86 \omega$ and  
$\phi \approx 1.73 \pi$, respectively. It is found that $x_{t_1}$ becomes minimum for 
$\phi = 0.75 \pi$. Now, because of the proportionality $ T \propto {\text{erf}}( x_{t_1} / \sqrt{2}
\sigma(t_1) )$, the maximum of $T$ coincides with the maximum of $x_{t_1}$. This explains the 
location of resonances of the transmission probabilities.

\section{Conclusions}

The study of dissipative tunnelling by an inverted parabolic barrier carried out here
clearly shows how the decoherence process is increasing gradually with the dynamical 
regime considered and governed by $\epsilon$ as well as with the friction and field parameters.
However, the important point is that when comparing the Kostin and CK approaches, the 
tunnelling probabilities are different as also observed by Tokieda and Hagino 
\cite{ToHa-PRC-2017}. This discrepancy comes from the differential equation 
governing the width of the Gaussian wave packet where both approaches differ. At this point, it
is difficult to discern which approach is better suited when a comparison with experimental 
results is carried out. We lean towards the Kostin approach due to, at least, two points: 
(i) When an interaction with an environment (bath, measuring apparatus, etc) is present, 
linear quantum mechanics is no longer applicable and nonlinear differential equations have to 
be implemented for a proper description of the corresponding open quantum dynamics, and (ii)
the nonlinear Kostin equation comes from the standard Langevin equation which is also issued from
a Caldeira-Leggett Hamiltonian formalism, whereas the CK approach is seen more like 
a phenomenological or effective one. In any case, new theoretical developments and numerical 
simulations are necessary to be implemented and compared with existing experimental results in order to have
a better description of this open dynamics. A natural extension of this work is to include the stochasticity
into the dynamics through a random force or noise term.

\vspace{2cm}
\noindent
{\bf Acknowledgements}
SVM acknowledges partial support from the University of Qom and SMA support from 
the Ministerio de Econom\'ia y Competitividad (Spain) under the Project 
FIS2014-52172-C2-1-P.

%=====================================================
%=====================================================
%********************************************************

%===========================
%+++++++++++++++++++++++++++++++++++++++++++++++++++
%+++++++++++++++++++++++++++++++++++++++++++++++++++
%
\end{document}